\documentstyle[aps,prl,psfig,amsfonts,amssymb,cite,multicol]{revtex}

\newcommand{\tie}[1]{#1_{\mbox{\tiny e}}}
\newcommand{\tih}[1]{#1_{\mbox{\tiny h}}}
\newcommand{\tiel}[1]{\overline{#1}_{\mbox{\tiny e}}}
\newcommand{\tihl}[1]{\overline{#1}_{\mbox{\tiny h}}}

\newcommand{\sie}{_{\mbox{\tiny e}}}
\newcommand{\sih}{_{\mbox{\tiny h}}}
\newcommand{\sio}{_{\mbox{\tiny o}}}

\newcommand{\uie}{^{\mbox{\tiny e}}}
\newcommand{\uih}{^{\mbox{\tiny h}}}

\newcommand{\ind}[1]{\mbox{\tiny #1}}

\begin{document}

\title{Optical properties of quantum wires:\\
Disorder--scattering in the Lloyd--model}
\author{Christian Fuchs and Ralph v.~Baltz}
\address{Institut f\"ur Theorie der Kondensierten Materie,
         76128 Karlsruhe, Germany}
\date{\today}
\maketitle

\begin{abstract}
The Lloyd model is extended to the exciton problem in
quasi one--dimensional structures to study the interplay between
the Coulomb attraction and disorder scattering.
Within this model the averaging and resummation of the 
locator series can be performed analytically.
As an application, the optical absorption in quantum box wires is
investigated. Without electron--hole interaction, 
fluctuations in the well--width lead to an 
asymmetric broadening of the minibands with respect to the lower and
upper band--edges. 
\end{abstract}         

\begin{multicols}{2}

\section{Introduction}
The optical properties of semiconductors near the band--edge 
are substantially influenced by the attractive electron--hole interaction 
which leads to pronounced excitonic lines
and enhanced absorption above the band gap\cite{Haug-K}.
These phenomena become even more pronounced in systems
of reduced effective dimensionality such as quantum--wells, wires or 
dots.
In addition, the influence of disorder and impurity scattering 
increases with decreasing dimension.
The latter phenomenon, however, is considered usually phenomenologically
by replacing the light frequency $\omega$ by $\omega+i\gamma$, 
or convoluting the  spectrum with an appropriate smoothing function of
width $\gamma$.

In this article, we investigate the interplay between excitonic effects and
disorder--scattering in quasi one--dimensional structures, in particular
quantum wires which
possess a periodic step--like modulation along the one--dimensional direction, 
(quantum box structures, QBS),  Fig.~1.
Such structures have been proposed as favorite candidates to generate 
Bloch oscillations \cite{Sakaki,Noguchi,LGeller},
other examples are  quasi one--dimensional molecular crystals
\cite{Abe,Ogawa,Ishi,Ando,Kim},
yet no detailed theoretical study of the optical properties has been 
published.
The optical properties of one--dimensional systems are exceptional
as even a small attractive electron--hole interaction leads to a drastic
reduction in the oscillator strength at the gap energy so that the 
identification of the gap energy from optical absorption data is nontrivial.

Our paper is organized as follows: Sect. II summarizes the basic
description of the optical susceptibility in terms of the resolvent operator.
In Sects. III and IV the averaging on the disordered chain is performed.
Within a Wannier representation the averaging and resummation of the 
locator expansion  can be done exactly, provided 
(a) only the lowest electron/hole subbands are considered, and 
(b) a Lorentzian probability
distribution for the disorder averaging is used. The latter is known as the
Lloyd--model\cite{Lloyd} which has been first used in connection with
the calculation of the density of states in a disorded metal.
Due to the simplicity in calculating various averages this model has been 
found useful by many other researchers, in particular Refs.%
\cite{Hirota,John,Hoshino,Luttinger,Greer,Kivelson,Lu,Cohen,Simon,
Johnston,Thouless,MacK,Rod1,Rod2,Mudry}.
However, apart from Hoshino's work\cite{Hoshino} 
on the dc conductivity (CPA, neglecting vertex corrections)
all previous applications of the Lloyd-model are 
exclusively restricted to single--particle quantities.
Our work extends the Lloyd--model to a two--particle quantity
and we give an exact solution for the optical absorption within a two--band 
model. Sect. V gives some analytical and numerical results for QBS near the 
fundamental gap and Sect. VI contains our conclusions.

\begin{figure}
\centerline{\psfig{file=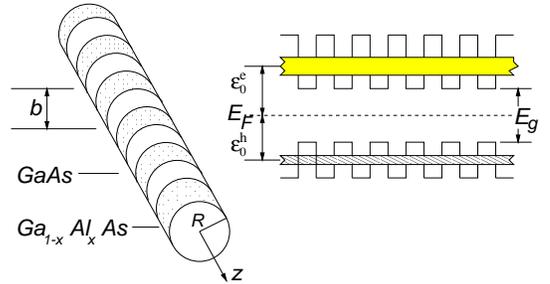,width=7cm}}

\medskip
\caption{ Quasi--one--dimensional quantum--box--structure (left) 
and sketch of lowest valence/conduction subbands (right). (Not to scale).}
\end{figure}

\section{Basics}
Our starting point is the excitonic contribution to the optical
suszeptibility $\chi(\omega)$ in a two--band effective mass 
approximation\cite{Haug-K,Glutsch} 
\begin{equation}
 \chi(\omega)=-\frac{2|d\ind{cv}|^2}{\epsilon_0 \Omega} 
               \sum_{n}{ \frac{1}{\hbar \omega + \imath \delta - E_n}
          \left|\int_{\Omega} d^3\vec r \, \Psi_n(\vec r, \vec r)\right|^2 },
\label{chi}
\end{equation}
where the nonresonant part has been neglected, ($\omega>0$).  $\Omega$
is the volume of the sample, and
$|d_{cv}|^2$ is the dipole matrix element between the valence and 
conduction band. $\Psi_n(\vec r_e,\vec r_h)$, $E_n$ respectively 
denote the electron--hole 
envelope function and  energy, and $n$ labels discrete as
well as the continuous part of the excitonic spectrum. 
\begin{equation}
  \left[ H\sie + H\sih + V(\vec r\sie -\vec r\sih) \right]
  \Psi_n(\vec r\sie, \vec r\sih)  = E_n\Psi_n(\vec r\sie, \vec r\sih) .
 \end{equation}
$H_j$, j=e,h denote the electron/hole Hamiltonians and $V(\vec r)$
is the attractive electron--hole Coulomb potential. 
We assume a cylindrical wire of radius $R$ with an
infinite confining potential and restrict 
ourselves to the lowest electron/hole mini--bands with
wave functions  $\Psi_j(\vec r)=u(r_\perp)\psi_j(z)$.
$u(r_\perp)$ is radialsymmetric, nodeless and
$\psi_j(z)$ denote the wave function along the wire z--axis which obeys 
\begin{eqnarray}
  H_j\psi_j(z) &=& E_j \psi_j(z),\\
  H_j          &=& \frac{-\hbar^2}{2m_j}\frac{\partial^2}{\partial z^2} + 
                   V_j(z).
\end{eqnarray} 
$m_j$ are the effective electron/hole masses and $V_j(z)$ includes
the band offset as well as the localization  energy 
due to the radial confinement.
For a homogeneous wire the energy spectrum consists of  free electron like 
mini bands with parabolic dispersions. 
These bands will be additionally split into subbands by the periodic 
modulation in composition  along the wire axis.
In the following we restrict ourselves to the lowest electron and hole 
subbands which are conveniently described in terms of Wannier--functions
$a_j(z-pb)$ where $p$ labels sites and $b$ is the period of the box structure,
Fig.~1.
By an appropriate choice of the phases of the Bloch--functions $\psi_j(k,z)$
the Wannier--functions can be constructed in such a way that they are
real, symmetric, and exponentially localized\cite{Kohn}. 
Furthermore, the Wannier--functions of the lowest hole and electron 
subbands both have the same even parity.
\begin{eqnarray}
&\Psi_n({z}\sie, {z}\sih)& =
\sum_{p\sie,p\sih}
\psi^{(n)}_{p\sie,p\sih} a\sie(z\sie- p\sie b) a\sih(z\sih- p\sih b), 
\label{ansatz}\\
 &\sum_{\tiel{p},\tihl{p}}&{\cal H}_{p\sie p\sih \tiel{p} \tihl{p}}\, 
 \psi^{(n)}_{\tiel{p} \tihl{p}} = E_n \, \psi^{(n)}_{p\sie p\sih}. \label{bas3}
\end{eqnarray}
Within a next neighbour approximation the the matrix elements
of the exciton Hamiltonian 
${\boldmath\cal H} = {\boldmath\cal H}\uie + {\boldmath\cal H}\uih + 
{\boldmath\cal V}$
read 
\begin{eqnarray}
\label{lat-ham}
{\cal H}_{\tie{p}\tih{p}\tiel{p}\tihl{p} } &=&
\delta_{\tih{p}\tihl{p}}
{\cal H}\uie_{\tie{p}\tiel{p}} +
\delta_{\tie{p}\tiel{p}}
{\cal H}\uih_{\tih{p}\tihl{p}}
+
{\cal V}_{\tie{p}\tih{p}\tiel{p}\tihl{p}}, \label{offc1} \label{HamOp}\\
  {\cal H}^{(j)}_{p,\overline p}&=&{\cal D}^{(j)}_{p,\overline p}
 +{\cal N}^{(j)}_{p,\overline p}, \label{offc3}\\
  {\cal D}^{(j)}_{p,\overline p}&=&
   \delta_{p,\overline p} \epsilon_{\mbox{\tiny 0}}^{(j)},\\
  {\cal N}^{(j)}_{p,\overline p}&=&
    t_1^{(j)} \{ \delta_{\overline p,p +1}  +\delta_{p,\overline p +1} \}.
\end{eqnarray}
$\epsilon_{\mbox{\tiny 0}}^{(j)}, t_1^{(j)}$ respectively denote the 
on--site energies and transfer elements between neighbouring wells and
${\cal V}_{\tie{p}\tih{p}\tiel{p}\tihl{p}}\approx
V(\tie{p}-\tih{p}) \delta_{\tie{p}\tiel{p}} \delta_{\tih{p}\tihl{p}}$ 
is the electron--hole Coulomb matrix element.

To calculate the optical absorption we rewrite Eq.~(\ref{chi}) in terms 
of the resolvent operator  ${\boldmath\cal G}(\tilde E)$ 
\begin{eqnarray}\label{ichi}
 \chi(\hbar \,\omega) 
 &\propto& - \sum_{p,\overline p} 
       {\cal G}_{p,p,\overline p,\overline p}
 \left( \hbar\,\omega +\imath\delta \right)  , \label{bas4}\\
 {\mathbf\cal G} (\tilde E) &=& \left( \tilde E - {\mathbf\cal H}\right)^{-1} ,
 \label{bas5}
\end{eqnarray}
where $\tilde E=E+\imath\delta$, $\delta>0$.

Without disorder,
the electron--hole wave function 
obeys the Bloch theorem and, hence,  it can be labelled by a wave number 
$k=2\pi\kappa/N$, $\kappa=0,\pm 1,\dots \pm N/2$ 
\begin{equation}
 \psi^{(n,k)}_{p\sie +\nu, p\sih +\nu}= e^{\imath \, kb \, \nu}
 \psi^{(n,k)}_{p\sie, p\sih}.
 \label{bloch}
\end{equation}
$\nu=1,2\dots N$, where $N$ is the number of sites on the chain. 
Periodic boundary conditions are implied. Hence,
\end{multicols}
\begin{equation}
 {\cal G}_{p\sie, p\sih, \overline p\sie, \overline p\sih} (\tilde E)=
 \sum_{n,k} 
 \frac{ \psi^{*(n,k)}_{p\sie,p\sih} 
        \psi^{(n,k)}_{\overline p\sie, \overline p\sih} }
      {\tilde E - E_{n,k} }
 ={\cal G}_{p\sie +\nu, p\sih +\nu, 
                    \overline p\sie +\nu, 
                    \overline p\sih +\nu}(\tilde E).
\end{equation}
Furthermore, only $k=0$ states 
$\psi^{(n,k=0)}_{p\sie,p\sih}=\phi^{(n)}_{p\sie - p\sih}$ 
contribute to the optical absorption 
\begin{equation}
  \sum_{p,\overline p} { {\cal G}_{p, p, \overline p, \overline p} (\tilde E) } 
    =
N \sum_p { {\cal G}_{p,p,0,0} (\tilde E) } 
    = 
N \sum_n { \frac{ \phi^{*(n)}_0 \phi^{(n)}_0 } {\tilde E - E_n } }
    = N G_{00}(\tilde E).
\end{equation}
The $k=0$ states are translational invariant 
and effectively describe a single particle in an external potential
\begin{equation}
 \left( t_1^{\ind{h}} +  t_1^{\ind{e}}\right)
 \left( \phi^{(n)}_{r+1} +  \phi^{(n)}_{r-1}\right)
 +
 \left(
  \epsilon_0^{\ind{h}} +  \epsilon_0^{\ind{e}}
 \right) \phi^{(n)}_{r} +  V_r \phi^{(n)}_r
 = E_n \phi^{(n)}_r,\quad r\in {\cal Z}.
\end{equation}
\begin{multicols}{2}
$G_{r,\bar r}(\tilde E)$ is the corresponding Green--function which obeys
the Dyson--equation
\begin{equation}
  {\mathbf G} = {\mathbf G}_0 + {\mathbf G}_0{\mathbf V G} . 
\label{Dyson}
\end{equation}
Boldface symbols denote matrices,
${\mathbf G}_0$ is the Green--function for $V_r=0$, and
$\mathbf V$ is a diagonal
matrix, $V_{rr'}=V_r\delta_{rr'}$.
The formal solution of Eq.~(\ref{Dyson}) is obtained by matrix inversion 
\begin{equation}
  {\mathbf G} = ({\mathbf 1-G}_0{\mathbf V})^{-1}{\mathbf G}_0 .
\label{D-sol}
\end{equation}

In addition to the attractive electron-hole interaction 
we shall consider fluctuations in the on--site energies $\epsilon_0$ as well 
as in the transfer elements $t_1$ between adjacent wells of the 
Kronig--Penny potential. Such fluctuations arise from
compositional and structural disorder  of the wire and we assume that these
fluctuations preserve the radial symmetry  so that mixing of higher 
subbands will not be important.
\end{multicols}

\section{Barrier fluctuations}

Fluctuation in the barrier thickness lead to coupled fluctuations 
in the on--site energies and the transfer elements
between sites $q$ and $q+1$ (nondiagonal disorder).
In particular,  we consider a correlated linear change of the 
onsite energies as mentioned by John and Schreiber\cite{John}.
For the lowest subband $t_1$ is negative, thus,
corrections to the onsite energies and hopping element
have the same sign, i.e. $\epsilon_q=\alpha\Delta t$, $\alpha>0$.
(The coupling of diagonal and non--diagonal disorder will be essential
to guarantee the convergence of the locator expansion and non--negative
absorption).
For simplicity, fluctuations in the electron and hole parameters are assumed
to be independent so that the Hamiltonians Eq.~(\ref{lat-ham}) are 
replaced by
\begin{eqnarray}
  {\cal D}^{(j)}_{p,\overline p} &=& \epsilon_{\mbox{\tiny 0}}^{(j)}
          +\alpha^{(j)} \Delta t^{(j)}_{p,p +1}
          +\alpha^{(j)} \Delta t^{(j)}_{p,p -1} ,\\
  {\cal N}^{(j)}_{p,\overline p}&=&
   \delta_{\overline p,p+1}
   \left\{ t_1^{(j)} + \Delta t^{(j)}_{\overline p,p+1} \right\}
  +\delta_{p,\overline p +1}
   \left\{ t_1^{(j)} + \Delta t^{(j)}_{p,\overline p +1}
   \right\}.
\end{eqnarray}
In matrix notation
${\boldmath\cal H} = {\boldmath\cal D} + {\boldmath\cal N} + {\mathbf\cal V}$,
where ${\boldmath\cal D}, {\boldmath\cal N}$ include electron and 
hole parts. To perform the disorder averaging the locator 
expansion of the resolvent will be used\cite{Ziman}
\begin{eqnarray}
 {\boldmath\cal G}(\tilde E) &=&
 \sum_{n=0}^{\infty}
 {\boldmath\cal G}^{(n)}(\tilde E) ,\label{expa}\\
 {\boldmath\cal G}^{(n)}(\tilde E ) &=& 
 \frac{1}{\tilde E - {\boldmath\cal D}}
 \left[
 {\cal M}\frac{1}{\tilde E - {\boldmath\cal D}}
 \right]^{n} , \label{expaf}
\end{eqnarray}
where ${\boldmath\cal M} = {\boldmath\cal N} + {\boldmath\cal V}$.
As ${\boldmath\cal D}$ is a diagonal matrix, we  have  ($n\geq 2$):
\begin{eqnarray}
{\cal G}^{(0)}_{\tie{p}\tih{p}\tiel{p}\tihl{p}}(\tilde E) &=& 
\frac{\delta_{\tie{p}\tie{p}}\delta{\tihl{p}\tihl{p}}}
     {\tilde E - {\cal D}_{\tie{p}\tih{p}\tie{p}\tih{p}}}
\label{gnull},\\
{\cal G}^{(1)}_{\tie{p}\tih{p}\tiel{p}\tihl{p}}(\tilde E) &=& 
\frac{1}
     {\tilde E - {\cal D}_{\tie{p}\tih{p}\tie{p}\tih{p}}}
\frac{{\cal M}_{\tie{p}\tih{p}\tiel{p}\tihl{p}}}
     {\tilde E - {\cal D}_{\tiel{p}\tihl{p}\tiel{p}\tihl{p}}}
\label{geins},\\
 {\cal G}^{(n)}_{\tie{p}\tih{p}{\tiel{p}\tihl{p}}}
 (\tilde E)
 &=& \sum_{\tie{p}^1 \tih{p}^1}
   \cdots
   \sum_{\tie{p}^{(n-1)} \tih{p}^{(n-1)}}
 \frac{1 }
      {\tilde E - {\cal D}_{\tie{p}\tih{p}\tie{p}\tih{p}}}
 \frac{ {\cal M}_{\tie{p}\tih{p}\tie{p}^{(1)}\tih{p}^{(1)}} }
      {\tilde E - {\cal D}_{\tie{p}^{(1)}\tih{p}^{(1)}
                            \tie{p}^{(1)}\tih{p}^{(1)}}}
 \frac{ {\cal M}_{\tie{p}^{(1)}\tih{p}^{(1)}\tie{p}^{(2)}\tih{p}^{(2)}} }
      {\tilde E - {\cal D}_{\tie{p}^{(2)}\tih{p}^{(2)}
                            \tie{p}^{(2)}\tih{p}^{(2)}}}
 \cdots\nonumber\\
  &&
 \hspace{2cm}\cdots
 \frac{ {\cal M}_{\tie{p}^{(n-2)}\tih{p}^{(n-2)}
                  \tie{p}^{(n-1)}\tih{p}^{(n-1)}} }
      {\tilde E - {\cal D}_{\tie{p}^{(n-1)}\tih{p}^{(n-1)}
                            \tie{p}^{(n-1)}\tih{p}^{(n-1)}}}
 \frac{ {\cal M}_{\tie{p}^{(n-1)}\tih{p}^{(n-1)}\tiel{p}\tihl{p}} }
      {\tilde E - {\cal D}_{\tiel{p}\tihl{p}
                            \tiel{p}\tihl{p}}}.
\label{genn}
\end{eqnarray}

Next, the disorder averaging will be performed term by term. 
Concerning the factors
\begin{equation}
 \frac{ {\cal M}_{\tie{p}\tih{p}\tiel{p}\tihl{p}}}
      {\tilde E - {\cal D}_{\tiel{p}\tihl{p}\tiel{p}\tihl{p}}}
 \label{factor} 
\end{equation}
four cases must be distinguished:\\ 
{\bf 1:} ($\tie{p} = \tiel{p}$,  $\tih{p} = \tihl{p}$)
\begin{eqnarray}
 \frac{ {\cal M}_{\tie{p}\tih{p}\tiel{p}\tihl{p}}}
      {\tilde E - {\cal D}_{\tiel{p}\tihl{p}\tiel{p}\tihl{p}}}
 =
 \frac{{\cal V}_{\tiel{p}\tihl{p}\tiel{p}\tihl{p}}}
      {
       \tilde E - \epsilon_{\mbox{\tiny 0}}\uie 
                - \epsilon_{\mbox{\tiny 0}}\uih
       - \alpha\uie 
         \left\{
          \Delta t \uie_{\tiel{p},\tiel{p}+1}
          +
          \Delta t \uie_{\tiel{p}-1,\tiel{p}} 
         \right\}
       - \alpha\uih 
         \left\{
          \Delta t \uih_{\tihl{p},\tihl{p}+1}
          +
          \Delta t \uih_{\tihl{p}-1,\tihl{p}} 
         \right\}       
      }\label{case1}
\end{eqnarray}
{\bf 2:} ($\tie{p} = \tiel{p}$,  $\tih{p} \not= \tihl{p}$)
\begin{eqnarray}
 \frac{ {\cal M}_{\tie{p}\tih{p}\tiel{p}\tihl{p}}}
      {\tilde E - {\cal D}_{\tiel{p}\tihl{p}\tiel{p}\tihl{p}}}
 =
 \frac{
       {\cal V}_{\tie{p}\tih{p}\tie{p}\tihl{p}}
       +
       \delta_{\tihl{p},\tih{p}+1} 
        \left\{
         t_1\uih + \Delta t\uih_{\tih{p},\tih{p}+1}
        \right\}
       +
       \delta_{\tih{p},\tihl{p}+1} 
        \left\{
         t_1\uih + \Delta t\uih_{\tihl{p},\tihl{p}+1}
        \right\}
      }
      {
       \tilde E - \epsilon_{\mbox{\tiny 0}}\uie 
                - \epsilon_{\mbox{\tiny 0}}\uih
       - \alpha\uie 
         \left\{
          \Delta t \uie_{\tiel{p},\tiel{p}+1}
          +
          \Delta t \uie_{\tiel{p}-1,\tiel{p}} 
         \right\}
       - \alpha\uih 
         \left\{
          \Delta t \uih_{\tihl{p},\tihl{p}+1}
          +
          \Delta t \uih_{\tihl{p}-1,\tihl{p}} 
         \right\}       
      }\label{case2}
\end{eqnarray}
{\bf 3:} ($\tie{p} \not= \tiel{p}$,  $\tih{p} = \tihl{p}$)
\begin{eqnarray}
 \frac{ {\cal M}_{\tie{p}\tih{p}\tiel{p}\tihl{p}}}
      {\tilde E - {\cal D}_{\tiel{p}\tihl{p}\tiel{p}\tihl{p}}}
 =
 \frac{
       {\cal V}_{\tie{p}\tih{p}\tiel{p}\tih{p}}
       +
       \delta_{\tiel{p},\tie{p}+1} 
        \left\{
         t_1\uie + \Delta t\uie_{\tie{p},\tie{p}+1}
        \right\}
       +
       \delta_{\tie{p},\tiel{p}+1} 
        \left\{
         t_1\uie + \Delta t\uie_{\tiel{p},\tiel{p}+1}
        \right\}
      }
      {
       \tilde E - \epsilon_{\mbox{\tiny 0}}\uie 
                - \epsilon_{\mbox{\tiny 0}}\uih
       - \alpha\uie 
         \left\{
          \Delta t \uie_{\tiel{p},\tiel{p}+1}
          +
          \Delta t \uie_{\tiel{p}-1,\tiel{p}} 
         \right\}
       - \alpha\uih 
         \left\{
          \Delta t \uih_{\tihl{p},\tihl{p}+1}
          +
          \Delta t \uih_{\tihl{p}-1,\tihl{p}} 
         \right\}       
      }\label{case3}
\end{eqnarray}
{\bf 4:} ($\tie{p} \not= \tiel{p}$,  $\tih{p} \not= \tihl{p}$)
analogous to case 1.

The averaging procedure requires the evaluation of the following multiple 
integrals
\begin{eqnarray}
 \left\langle
 {\cal G}^{(n)}_{\tie{p}\tih{p}\tiel{p}\tihl{p}}(\tilde E)
 \right\rangle
 &=&
 \int_{-\infty}^{\infty}
 d \Delta t_{12}\uie \,
 \ldots
 d \Delta t_{N 1}\uie \,
 d \Delta t_{12}\uih \,
 \ldots
 d \Delta t_{N 1}\uih \,
 P(\Delta t_{12}\uie, \gamma\uie)
 \ldots
  P(\Delta t_{N 1}\uie, \gamma\uie) \times\nonumber \\
 &\times&
  P(\Delta t_{12}\uih, \gamma\uih)
 \ldots
  P(\Delta t_{N 1}\uih, \gamma\uih)    
 {\cal G}^{(n)}_{\tie{p}\tih{p}\tiel{p}\tihl{p}}
 (
 \tilde E, \Delta t_{12}\uie,
 \ldots
 \Delta t_{N 1}\uie ,
 \Delta t_{12}\uih ,
 \ldots
 \Delta t_{N 1}\uih  ) , \label{disgf}
\end{eqnarray}
\begin{multicols}{2}
where $P(\Delta,\gamma)$ is the probability distribution function on the
disorder configurations.  
In the Lloyd--model this function is a Lorentzian
\begin{equation}
  P(\Delta,\gamma) = \frac{\gamma}{\pi}\frac{1}{\Delta^2 + \gamma^2},
\label{lloyd}
\end{equation}
where $\gamma$ parametrizes the width of the distribution. 

The  integrations in Eq.~(\ref{disgf}) will be performed step by step
in the indicated sequence.
In Eqs.~(\ref{gnull}-\ref{genn}) each factor is a
holomorphic function of $\Delta t\uih_{12}$ in the lower
complex plane including the real axis, in addition, this function is
bounded if $\alpha\uih\not= 0$. 
Thus, the averaging can be performed by contour integration 
closing the integration path in the lower  
$\Delta t\uih_{12}$--plane where Eq.~(\ref{lloyd}) has a first order pole.
This just leads to a replacement of 
$\Delta t_{12}\uih$ by $-\imath\gamma\uih$. 
Further integrations follow the same reasoning.
Hence, the resummation of the locator series is trivially possible
and, as a result, we obtain
\begin{equation}
 \overline{\boldmath\cal G}:= 
 \left\langle {\boldmath\cal G} ({\tilde E}) \right \rangle
 =  \left( \tilde E - \overline {\boldmath\cal H}  \right)^{-1}.
 \label{inverse}
\end{equation}
${\boldmath\cal H}$ is an effective Hamiltonian where the
electron/hole parts of Eqs.~(\ref{lat-ham}) are replaced by
\begin{eqnarray}
 \overline{\cal H}_{p,\overline p}^{(j)} =
  &\delta_{p,\overline p}&
  \left(
   \epsilon^{(j)}_{\mbox{\tiny o}} - 2 \imath \gamma^{(j)} \alpha^{(j)}
  \right)\nonumber\\
  &+&
  \left(
   \delta_{p,\overline p +1} + \delta_{\overline p, p+1} 
  \right)
  \left(
   t_1^{(j)} - \imath\gamma^{(j)}
  \right) .
\label{G-corrdis}
\end{eqnarray}
Thus, the influence of barrier--disorder is captured by the resolvent
of the pure wire where
$\epsilon^{(j)}_{\mbox{\tiny o}}, t_1^{(j)}$ are replaced by complex 
parameters 
${\bar\epsilon}^{(j)}_{\mbox{\tiny o}}=
\epsilon^{(j)}_{\mbox{\tiny o}}-2\imath\gamma^{(j)}\alpha^{(j)}$, 
${\bar t}_1^{(j)}=t_1^{(j)}- \imath\gamma^{(j)}$.
Although $\overline{\cal H}$ is no longer hermitian 
eigenstates exist and, moreover, these are  of the same form as
for the pure wire, in particular the eigenvalues are
\begin{equation}
E^{(j)}(k)={\bar\epsilon}^{(j)}_{\mbox{\tiny o}} + 2{\bar t}_1^{(j)}\cos (kb).
\end{equation}
For $t_1^{(j)}<0$ the lower/upper band edges are at $kb=0/\pi$
so that the complex electron/hole energies are  distributed between 
$\epsilon_{\mbox{\tiny o}}-2|t_1|-2\imath\gamma(\alpha+1)$ and
$\epsilon_{\mbox{\tiny o}}+2|t_1|-2\imath\gamma(\alpha-1)$.
The parameter $\alpha$ must be restricted to  $\alpha\geq 1$ for 
both electrons and holes, otherwise there will be energies
with a positive imaginary part,
i.e. the Green--function will have poles in the upper energy plane
which would violate causality. 
In addition,  the optical absorption 
would  become negative in some part of the spectrum.
Note, that the correlated fluctuations in the diagonal/nondiagonal
disorder will not simply lead to a Lorentzian smoothening of the 
absorption spectrum. In particular, for $\alpha=1$ the correlated shift 
of the on-site 
energies and transfer elements exactly cancel at the upper band edge 
so that the square root singularity of the density of states survives
if electron--hole interaction is omitted. 
\end{multicols}

\section{Well fluctuations}

Fluctuations in the well sections predominantly lead to fluctuations
in the onsite energies $\epsilon_q$. For convenience, fluctuations in 
the hopping elements will be omitted,
however, a correlation of the electron/hole energies will be retained:
$\Delta_p\uih= \alpha \Delta_p\uie$, 
$\alpha\approx m_e/m_h >0$ (diagonal disorder).

To perform the disorder averaging a regrouping
of the Hamitonian as
${\boldmath\cal H}={\boldmath\cal B}+{\boldmath\cal M}$
is necessary, where
\begin{eqnarray}
 {\cal M}_{p_{\ind{e}} 
           p_{\ind{h}} 
           \overline p_{\ind{e}}
           \overline p_{\ind{h}}}
 &=&
 \delta_{\overline p_{\ind{h}} p_{\ind{h}}}
 \left\{
  \delta_{\overline p_{\ind{e}} p_{\ind{e}}}
  \epsilon\sio^{\ind{e}}
  +
  t_1^{\ind{e}}
  \left( \delta_{\overline p_{\ind{e}}, p_{\ind{e}}+1}
  +
  \delta_{\overline p_{\ind{e}}, p_{\ind{e}}-1}  
  \right) 
 \right\}
 \nonumber\\
 &+&
 \delta_{\overline p_{\ind{e}} p_{\ind{e}}}
 \left\{
  \delta_{\overline p_{\ind{h}} p_{\ind{h}}}
  \epsilon\sio^{\ind{h}}
  +
  t_1^{\ind{h}}
  \left( \delta_{\overline p_{\ind{h}}, p_{\ind{h}}+1}
  +
  \delta_{\overline p_{\ind{h}}, p_{\ind{h}}-1}  
  \right) 
 \right\} + {\cal V}_{\tie{p}\tih{p}\tiel{p}\tihl{p}},\\
 {\cal B}_{p_{\ind{e}}p_{\ind{h}} 
           \overline p_{\ind{e}}\overline p_{\ind{h}}}
 &=&
  \delta_{\overline p_{\ind{h}} p_{\ind{h}}}
  \Delta_{p_{\ind{h}}}\uih
  +
  \delta_{\overline p_{\ind{e}} p_{\ind{e}}}
  \Delta_{p_{\ind{e}}}\uie.
\end{eqnarray}

Using the locator expansion Eqs.~(\ref{expa},\ref{expaf})
with ${\boldmath\cal D}$ replaced by $\boldmath\cal B$   
the matrix elements of the resolvent become  ($n\geq 2$)
\begin{eqnarray}
 {\cal G}^{(0)}_{\tie{p}\tiel{p}}\delta_{\tih{p}\tihl{p}}
 &=&
 \frac{\delta_{\tie{p}\tih{p}\tiel{p}\tihl{p}}}
      {\tilde E - (\Delta_{\tie{p}} + \alpha\Delta_{\tih{p}}) }
 \label{g1}\\
 {\cal G}^{(1)}_{\tie{p}\tih{p}\tiel{p}\tihl{p}}
 &=&
  \frac{1}
      {\tilde E - (\Delta_{\tie{p}} + \alpha\Delta_{\tih{p}}) }
  \frac{ {\cal M}_{\tie{p}\tih{p}\tiel{p}\tihl{p}} }
      {\tilde E - (\Delta_{\tiel{p}} + \alpha\Delta_{\tihl{p}}) }
 \label{g2} \\   
 {\cal G}^{(n)}_{p_{\ind{e}}^1, p_{\ind{h}}^1, 
                 \overline p_{\ind{e}}, \overline p_{\ind{h}}}
 &=&
 \sum_{p_{\ind{e}}^{1}, p_{\ind{h}}^{1}}
 \cdots
 \sum_{p_{\ind{e}}^{n-1}, p_{\ind{h}}^{n-1}} 
 \frac{
         1
         }
        {\tilde E - (\Delta_{p_{\ind{e}}} + \alpha
         \Delta_{p_{\ind{h}}})}
 \frac{
         {\cal M}_{p_{\ind{e}} p_{\ind{h}} p_{\ind{e}}^1 p_{\ind{h}}^1 } 
       }
      {\tilde E - (\Delta_{p_{\ind{e}}^1} + \alpha
       \Delta_{p_{\ind{h}}^1})}
 \frac{
       {\cal M}_{p_{\ind{e}}^1 p_{\ind{h}}^1 p_{\ind{e}}^2 p_{\ind{h}}^2 }
       }
      {\tilde E - (\Delta_{p_{\ind{e}}^2} + \alpha
        \Delta_{p_{\ind{h}}^2})}
 \cdots \nonumber\\
&&\cdots \frac{
       {\cal M}_{p_{\ind{e}}^{n-2} p_{\ind{h}}^{n-2} 
                 p_{\ind{e}}^{n-1} p_{\ind{h}}^{n-1} }
      }
      {\tilde E - (\Delta_{p_{\ind{e}}^{n-1}} 
                   + 
                   \alpha
                   \Delta_{p_{\ind{h}}^{n-1}})}
\frac{
       {\cal M}_{p_{\ind{e}}^{n-1} 
                 p_{\ind{h}}^{n-1} 
                 \overline p_{\ind{e}} 
                 \overline p_{\ind{h}}}  
     }
 {\tilde E - 
 (\Delta_{\overline p_{\ind{e}}}
 + 
 \alpha
  \Delta_{\overline p_{\ind{h}}} )} . \label{g3}
\end{eqnarray}
The disorder averaging of $\cal G$ is done along the same route as before
\begin{equation}
 \overline{\boldmath\cal G}:=
 \int_{-\infty}^{\infty}
 d \Delta_{1} \,
 \cdots
 d \Delta_{N} \,
 \,P(\Delta_{1},\gamma)\,\cdots \,P(\Delta_{N},\gamma )
 \,{\cal G}^{(n)}_{\tie{p}\tih{p}\tiel{p}\tihl{p}}
 (
 \tilde E, \Delta_{1},
 \ldots,
 \Delta_{N}, ) . 
\label{disgf2}
\end{equation}

As a result, we obtain
\begin{equation}
{\bar {\boldmath\cal G}}(\tilde E):=
\left\langle {\boldmath\cal G}(\tilde E) \right\rangle 
         =\left( 
           E+\imath (1+\alpha)\gamma-{\boldmath\cal M} 
          \right)^{-1}.
\label{G-diagdis}
\end{equation}
Thus, with respect to the pure system, diagonal disorder just leads to
the replacement of $E=\hbar\omega$ by the complex
quantity $E\to E+\imath (1+\alpha)\gamma$, which 
leads to the expected Lorentzian broadening
of the optical absorption spectrum.

\begin{multicols}{2}
\section{Applications}

The Green--function 
of the noninteracting pure wire is well known from textbooks, 
e.g. Economou\cite{Economou}. In a sligthly rewritten form 
which is suitable for the analytic continuation to complex 
$T_0 = {\bar\epsilon}\uie\sio + {\bar\epsilon}\uih\sio$,
$T_1 = {\bar t}_1\uie + {\bar t}_1\uih$, ($\Re T_1<0$), 
it reads
\begin{equation}
 G^{(0)}_{lm}(\tilde E) = \frac{\rho_1^{|l-m|}(\tilde E)}
                         {\sqrt{({\tilde E}-T_0)^2 - 4T_1^2}}\, ,
\end{equation}
where
\begin{equation}
 \rho_1(\tilde E)=\left(
                  \tilde E - T_0 -\sqrt{(\tilde E -T_0)^2 - 4T_1^2}
                  \right)/(2T_1) .
\end{equation}
$\sqrt{x}$ denotes the square root whose imaginary part has the
same sign as $\Im x$, see Eqs.~(\ref{G-corrdis},\ref{G-diagdis}). 

First, we consider a local electron--hole interaction, $V_r=V_0\delta_{r,0}$,
$V_0<0$.
In this case  Eq.~(\ref{Dyson}) can be solved analytically
\begin{equation}
  G_{00}(\tilde E) = \frac{G^0_{00}(\tilde E)}{1-V_0\,G^0_{00}(\tilde E)}\,.
\end{equation}
Without disorder (i.e. real $T_0,T_1$) the optical absorption $\chi_2=\Im \chi$
of a wire becomes
\begin{eqnarray}
 \chi_2 ( \hbar \omega ) &=& \frac{|W|^2}{2\pi |T_{\ind{1}}|}
 \frac{\sqrt{1-{\omega'}^2}}
      {v_0^2  + 1 - {\omega' }^2 },
 \quad |{\omega'}| < 1,\\
 \chi_2 (\hbar \omega ) &=&\frac{ |W|^2}{2|T_{\ind{1}}|}
          \frac{v_0}{\sqrt{1 + v_0^2}}\;\delta 
           \left( {\omega' } +\sqrt{1 + v_0^2} \,\right),\\
       &&\hspace*{3.6cm} \omega'<-1 .\nonumber
\end{eqnarray}
where ${\omega'}=(\hbar\omega - T_0)/(2|T_1|)$, $v_0=|V_0/(2T_1)|$.
$W$ containes the interband dipole matrix element and an overlap integral  
between Wannier--functions.
A similar result has been obtained by Ishida et al.\cite{Ishi}.
Note, $\hbar\omega$ is measured with respect to the gap energy $E\ind{g}$ of the 
homogeneous wire, Fig.~1.

For a quantum wire we have to replace the three--dimensional
Coulomb potential by the envelope averaged potential
which can be approximated quite well  by\cite{Haug-K}
\begin{equation}
 V(z)=-\frac{e^2}{4\pi\epsilon_0{\bar\epsilon}}
       \frac{1}{|z|+\beta R},
\label{V1d}
\end{equation}
where R is the wire radius, $\beta\approx 0.3$, and
$\bar\epsilon$ is an average dielectric constant.
For a $GaAs$ based wire of radius $R=50$\AA,   
$V(0)\approx -0.1$eV which is approximately the full
electron--hole transition bandwidth $4|T_1|$, hence $v_0\approx 0.5$.  

To solve the Dyson equation Eq.~(\ref{Dyson}) the effective
exciton potential Eq.~(\ref{V1d}) 
is truncated to a finite range, $V_r=V(r b)=0$, $|r|> s$. 
Then, the optical absorption, Eq.(\ref{bas5}), is obtained numerically 
by solving a $(2s+1)$ dimensional linear set of equations for
$G_{l0}(\tilde E)$, $l=-s,\dots s$.
Figs.~2-6 display various examples for the perfect and disordered wire.

\section{Conclusions}

The optical absorption in a quasi onedimensional system sensitivly 
depends on both the electron--hole  interaction  and the disorder scattering.
Diagonal disorder just results in a Lorentzian
smearing of the optical lines, whereas correlated diagonal/offdiagonal 
disorder is different: the broadening is stronger at the lower than at 
the upper band edge, Fig.~2. 
However, the shape of the absorption spectrum is drastically
changed even by  a tiny electron--hole interaction:
The divergences at the band edges are
completely removed so that the identification of the gap is a nontrivial problem
in quasi one--dimensional semiconductors, Figs.~3-5.
For a local electron--hole interaction there is only but
a single exciton bound state, Fig.~3, whereas 
the (truncated) Coulomb potential leads to an additional  pronounced spike
near the gap and an oscillatory structure within the absorption band, Fig.~6.
These structures are insensitive with respect to the type of disorder.

\vfill
\begin{figure}
\centerline{\psfig{file=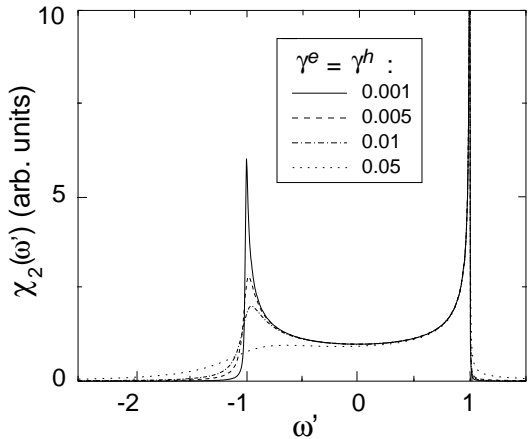,width=7cm}}

\medskip
\caption{Optical absorption of the noninteracting
             electron--hole system with varying barrier--disorder.\\
             $\alpha\uie = \alpha\uih = 1.1$.
             Energies are given in units of one half of the transtion band width,
             lower band edge is at $\omega'=-1$.}
\end{figure}

\begin{figure}
\centerline{\psfig{file=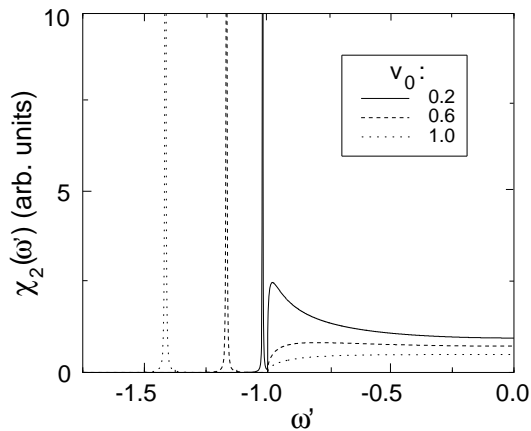,width=7cm}}

\medskip
\caption{Optical absorption of the pure wire
             with variation of the local electron--hole interaction.
             Energy scales as in Fig.~2.
             \vspace*{8pt}}
\end{figure}

\begin{figure}
\centerline{\psfig{file=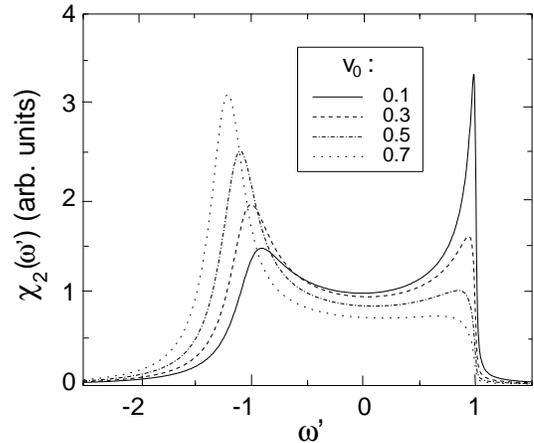,width=7cm}}

\medskip
\caption{Optical absorption with varying
              electron--hole interaction. Fixed barrier--disorder,
              $\alpha\uie = \alpha\uih = 1.1$,
              $\gamma\uie = \gamma\uih = 0.025$. 
              Energy scales as in Fig.~2. 
             \vspace*{-2pt}}
\end{figure}

\begin{figure}
\centerline{\psfig{file=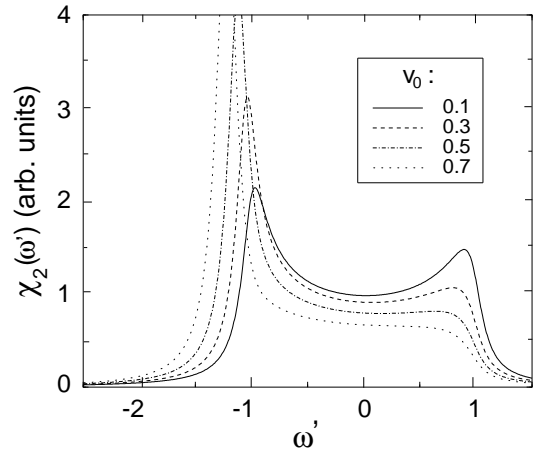,width=7cm}}

\medskip
\caption{Optical absorption
              with varying electron--hole interaction. Fixed well--disorder, 
              $(1+\alpha)\gamma=0.1$. Energy scales as in Fig.2.
              \vspace*{4pt}}
\end{figure}

\begin{figure}
\centerline{\psfig{file=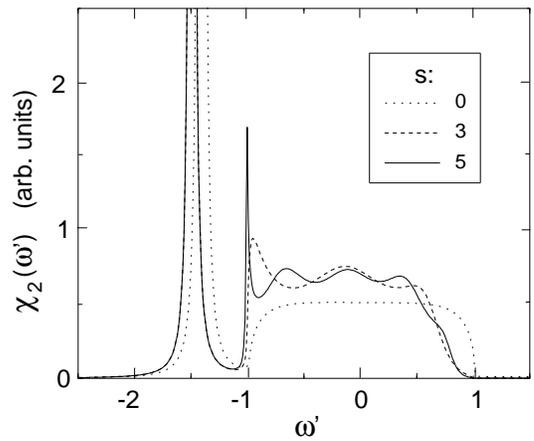,width=7cm}}

\medskip
\caption{Optical absorption with varying range $s$ of the truncated
             electron--hole Coulomb interaction, $v_0=1$. 
             Fixed well--disorder.$\alpha\uie=\alpha\uih=1.25$, 
             $\gamma\uie=\gamma\uih=0.005$.  
             Energy scales as in Fig.~2.}
\end{figure}

Our approach is based on the locator expansion of the Green--function 
(v.Neumann series) whose convergence is  rarely established in general. 
A more rigorous treatment, however,  shows that Eq.(\ref{G-corrdis}) 
is indeed correct if $\alpha>1$\cite{Fuchs}. For $0<\alpha<1$,
Eq.(\ref{G-corrdis}) leads to a negative absorption in some part of the 
spectrum which is definitively incorrect. It seems that the Lloyd model
cannot be formulated for purely nondiagonal disorder, i.e. $\alpha=0$. 
A similar problem is known
for the extension of the CPA to nondiagonal disorder\cite{CPA} , but
it may be also due to the pathological Lorentzian probability distribution,
Eq.~(\ref{lloyd}), of the Lloyd--model.

\section{Acknowlegdements}

We thank Profs. A. Schmid ($\dagger$) and Y. Levinson for many valuable
discussions and advice  on impurity problems. Part of this work was 
supported by the Deutsche Forschungsgemeinschaft through SFB195.


\end{multicols}

\end{document}